\documentclass[prl,twocolumn,superscriptaddress,amsmath,amssymb,aps,footinbib,10pt,nobibnotes]{revtex4-2}

\usepackage[utf8]{inputenc}
\usepackage{amsmath,amsthm,amssymb,amsfonts,braket}
\usepackage{placeins}[verbose]
\usepackage{tabularx, longtable, multirow}
\usepackage{graphicx}
\usepackage{textcomp,gensymb} 
\usepackage{dcolumn}
\usepackage{bm,dsfont}
\usepackage[english]{babel}
\usepackage{comment}
\usepackage[dvipsnames]{xcolor}
\usepackage{siunitx}
\usepackage[normalem]{ulem}
\usepackage[T1]{fontenc}
\usepackage{physics}
\usepackage{mathtools}
\usepackage[version=4,arrows=pgf-filled,textfontname=sffamily,
mathfontname=mathsf]{mhchem}

\renewcommand{\Re}{\mathrm{Re}}
\newcommand{\sinc}{\mathrm{sinc}}
\DeclareMathOperator{\sign}{sign}
\DeclareSIUnit\angstrom{\text {Å}}

\providecommand{\ignore}[1]{}

\newcommand{\more}[1]{{\color{Red}[\textbf{...}]}}

\begin{document}

\title{Topological shaping of vortex neutron beams using forked phase gratings}

\author{S. McKay}
\altaffiliation{Current address: Neutron Sciences Directorate, Oak Ridge National Laboratory, Oak Ridge, TN, 37830, USA}
\email{mckaysr@ornl.gov}
\affiliation{Department of Physics, Indiana University, Bloomington IN 47405, USA}
\affiliation{Center for Exploration of Energy and Matter, Indiana University, Bloomington, 47408, USA}
\author{S. R. Parnell}
\affiliation{ISIS, Rutherford Appleton Laboratory, Chilton, Oxfordshire, OX11 0QX, UK}
\author{R. M. Dalgliesh}
\affiliation{ISIS, Rutherford Appleton Laboratory, Chilton, Oxfordshire, OX11 0QX, UK}
\author{N. V. Lavrik}
\affiliation{Center for Nanophase Materials Sciences, Oak Ridge National Laboratory, Oak Ridge, TN 37831, USA}
\author{I. I. Kravchenko}
\affiliation{Center for Nanophase Materials Sciences, Oak Ridge National Laboratory, Oak Ridge, TN 37831, USA}
\author{Q. Le Thien}
\affiliation{Department of Physics, Indiana University, Bloomington IN 47405, USA}
\author{D. V. Baxter}
\affiliation{Department of Physics, Indiana University, Bloomington IN 47405, USA}
\affiliation{Center for Exploration of Energy and Matter, Indiana University, Bloomington, 47408, USA}
\affiliation{Quantum Science and Engineering Center, Indiana University, Bloomington, IN 47408, USA}
\author{G. Ortiz}
\affiliation{Department of Physics, Indiana University, Bloomington IN 47405, USA}
\affiliation{Quantum Science and Engineering Center, Indiana University, Bloomington, IN 47408, USA}
\author{R. Pynn}
\affiliation{Department of Physics, Indiana University, Bloomington IN 47405, USA}
\affiliation{Center for Exploration of Energy and Matter, Indiana University, Bloomington, 47408, USA}
\affiliation{Quantum Science and Engineering Center, Indiana University, Bloomington, IN 47408, USA}
\affiliation{Neutron Sciences Directorate, Oak Ridge National Laboratory, Oak Ridge, TN, 37830, USA}

\date{\today}

\begin{abstract}
Beams of light or matter that carry well-defined states of orbital angular momentum (OAM) are promising probes of topological and textured condensed matter systems such as magnetic skyrmions.
Using spin-echo small-angle neutron scattering (SESANS), we demonstrate the production of vortex neutron beams from forked phase gratings of various topological charges. In contrast to some previous techniques used to verify OAM production, SESANS is a more precise measurement of the neutron's OAM as it is a phase-sensitive, interferometric technique that directly measures the phase between the scattered neutron spin states.
\end{abstract}

\maketitle

\emph{Introduction.}---
Orbital angular momentum (OAM) beams of light or matter display a number of unique properties that make them promising candidates as probes of condensed matter systems \cite{Harris_2015,Shen_2019,Bliokh_2023}, such as \textit{superkicks}, enhanced scattering events from processes that are kinematically forbidden for beams without OAM \cite{Barnett_2013,Ivanov_2016,Afanasev_2021,Ivanov_2022,Li_2024}, and preferential absorption and scattering that depends on the probe's OAM state and the target's chirality \cite{Juchtmans_2015,VanVeenendaal_2015,Afanasev_2019,Forbes_2019,Bégin_2023}.
An OAM state is defined by its \textit{phase singularity} $e^{i \ell \phi}$, where $\ell \in \mathbb{Z}$ is the OAM quantum number and $\phi$ the azimuthal angle about the direction of travel. To preserve the single-valueness of the wavefunction, the intensity must fall to zero along the direction of travel, creating a phase-vortex around the intensity singularity.

\begin{figure}[t!]
    \centering
    \includegraphics[width=0.95\linewidth]{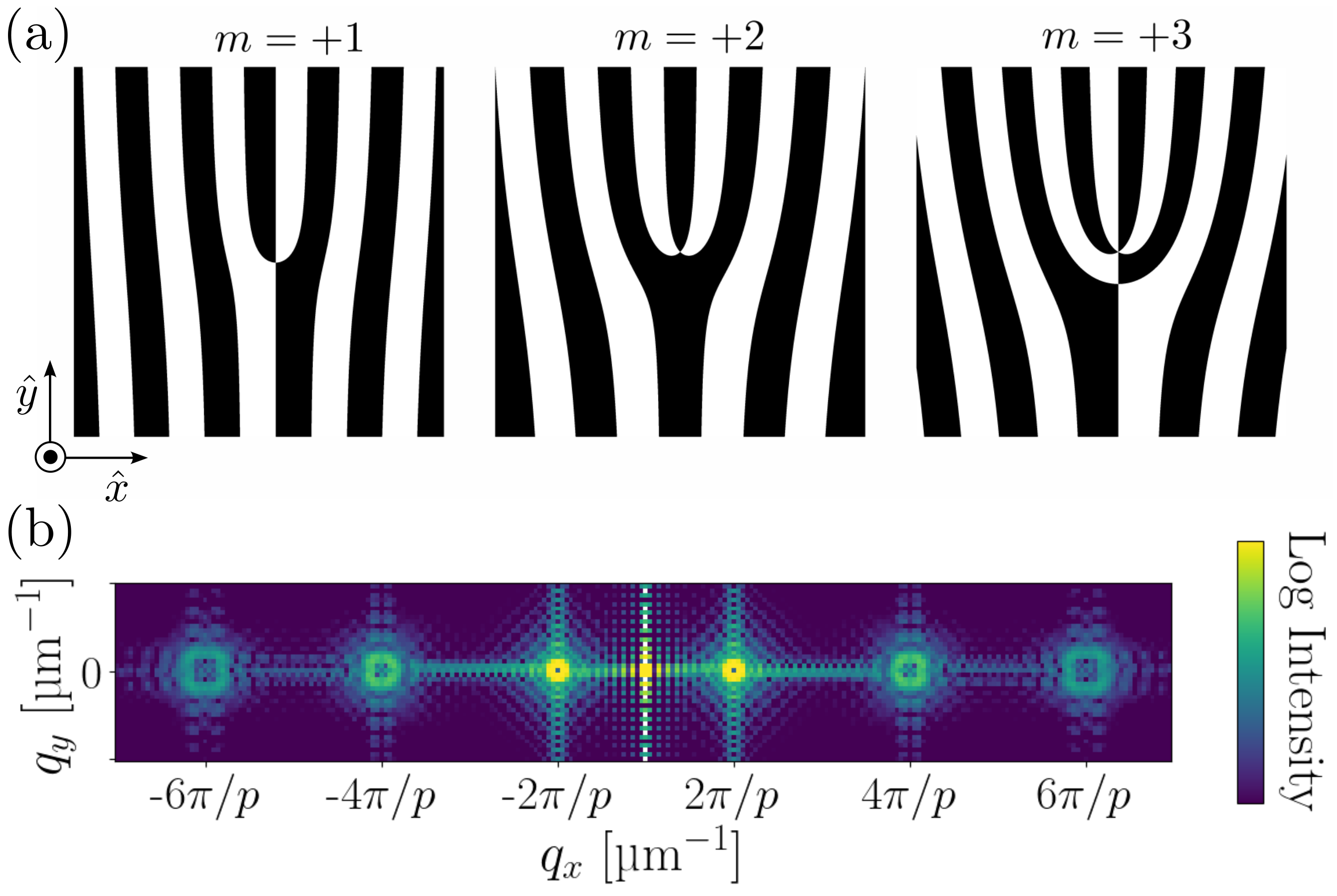}
    \caption{ \label{fig:FDGs and SANS}
    (a) Examples of binary forked phase grating profiles with topological charges $m =1,2,3$. (b) Plane wave simulation of the diffraction pattern produced from an $m=1$, period $p = \SI{2}{\micro \meter}$ forked phase grating vs. transverse momentum transfer $\bm q = (q_x,q_y)$.
    Notice that at the usual Bragg peak locations at $\bm q = (\pm 2 \pi n/p,0)$ for $n \neq 0$ we instead observe annular ``Bragg donuts,'' one of the characteristic indications of an OAM state. }
\end{figure}

\textit{Forked diffraction gratings} (FDGs) are often utilized as a robust method of OAM generation and measurement for both light \cite{Lee_2019,Ishii_2022,McCarter_2024} and matter waves \cite{verbeeck_2010,McMorran_2011,Saitoh_2013,Grillo_2014,Noguchi_2019}.
The characteristic profile of a FDG with grooves along the $\hat y$ direction is described by $\cos \alpha$ with $\alpha = (2\pi/p) x - m \phi$, where $p$ is the period of the grating, $m \in \mathbb{Z}$ the grating's \textit{topological charge}, and $\phi = \arctan(y/x)$.
The topological charge corresponds to the difference of the number of grooves above and below the topological defect at the grating's center as shown in Fig. \ref{fig:FDGs and SANS}(a). 
This function is the interference pattern of a plane wave and a charge $m$ OAM state, and so a FDG acts as a hologram, converting an incoming neutron without OAM into a conjugate pair of outgoing $\ell= \pm m$ OAM states \cite{Bazhenov_1990,Heckenberg_1992a,Heckenberg_1992b}.

Although the generation of OAM neutron beams has been reported using various techniques \cite{clark_2015,Sarenac_2016,Sarenac_2019,Sarenac_2022,Sarenac_2024,Geerits_2022,Geerits_2025,Geeritsb_2025}, it is generally difficult to unambiguously verify the production of neutron OAM \cite{Cappelletti_2018,Cappelletti_2021,Treimer_2025}, primarily due to the weak interactions of thermal and cold neutrons ($E \lesssim \SI{25}{\milli \electronvolt}$) with matter.
In contrast, here we demonstrate the production of neutron OAM from FDGs using \textit{spin-echo small-angle neutron scattering} (SESANS), which is a {\it phase-sensitive}, interferometric technique described in the next section.

\emph{Sample design and experiment.}---
Our silicon FDGs were etched using electron-beam lithography at the Center for Nanophase Materials Sciences at Oak Ridge National Laboratory (see supplemental material \cite{Supp}).
As our FDG grooves have a rectangular rather than sinusoidal profile, they also produce higher-order diffraction peaks that are weaker with larger OAM values.
For a 50\% duty cycle grating, the rectangular profile can be described by the indicator function 
\begin{equation} \label{eq:Rect FDG}
    \chi(x,y) = \frac{1}{2} \left[ 1 + \sign \left( \cos \alpha \right)\right].
\end{equation}
As the production of neutron OAM requires the FDG to be coherently illuminated, the size of the FDG must be on the order of the transverse \textit{intrinsic} coherence width $\Delta_t$ of the neutron, which we define to be the average transverse size of each of the mutually incoherent neutron wavepackets that form the total beam \footnote{The intrinsic coherence width $\Delta_t$ of a single neutron is distinct from the quantity we define as the transverse \textit{beam} coherence width $\beta_t$ of the ensemble of incoherent neutrons, which up to some numerical constants in the pinhole geometry is given by $\beta_t \sim \! \lambda L / a$ where $L$ is the propagation distance and $a$ the width of the source aperture. Under the assumption of coherent illumination of the source aperture (i.e., $\Delta_t \geq a$), we have $\beta_t \leq \Delta_t$.}.
Although measurements of $\Delta_t$ are experimentally challenging, most report $\Delta_t \gtrsim \! \SI{10}{\micro \meter}$ \cite{Shull_1969,Treimer_2006,Altissimo_2008,Wagh_2011,Majkrzak_2022}.
To enhance the signal to an observable level, the FDG motif must be repeated periodically as individual plaquettes over the area of the beam. Each of our plaquettes have a period $p=\SI{2}{\micro \meter}$, groove depth $d\approx 4$-$\SI{6}{\micro \meter}$, and total size of $10\times10$ \SI{}{\micro \meter^2}.
The total etched area of the wafer is $1\times1$ \SI{}{\centi \meter^2}.
As the region near the singularity is especially fragile, an additional small circular area was etched away near each plaquette's center; although we do not show this additional feature in Fig. \ref{fig:FDGs and SANS}(a), it is included in the simulations presented below.

\begin{figure*}[t!]
    \centering
    \includegraphics[width=0.95\linewidth]{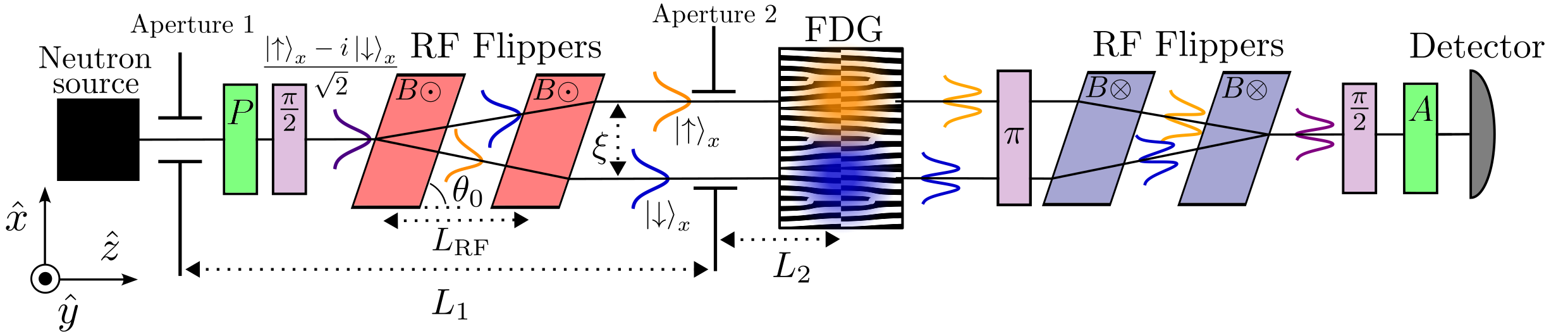}
    \caption{ \label{fig:SESANS setup}
    Experimental setup of the SESANS measurement.
    Collimation was performed by two square apertures placed before the polarizer (Aperture 1) of size $14 \times \SI{14}{\milli \meter^2}$ and before the FDG (Aperture 2) of size $6 \times \SI{6}{\milli \meter^2}$. The distances between the two apertures and the FDG and Aperture 2 are $L_1 = \SI{4.82}{\meter}$ and $L_2 = \SI{0.09}{\meter}$, respectively. The green components label ``P'' and ``A'' are the neutron spin polarizer and analyzer, respectively. The two $\pi/2$-flippers start and stop the neutron precession while the central $\pi$-flipper corrects for the magnetic inhomogeneities in the rf flippers.
    Additional weak guide fields are not shown. The wavepacket size and separation are exaggerated for clarity.}
\end{figure*}

\begin{figure}[ht!]
    \centering
    \includegraphics[width=0.99\linewidth]{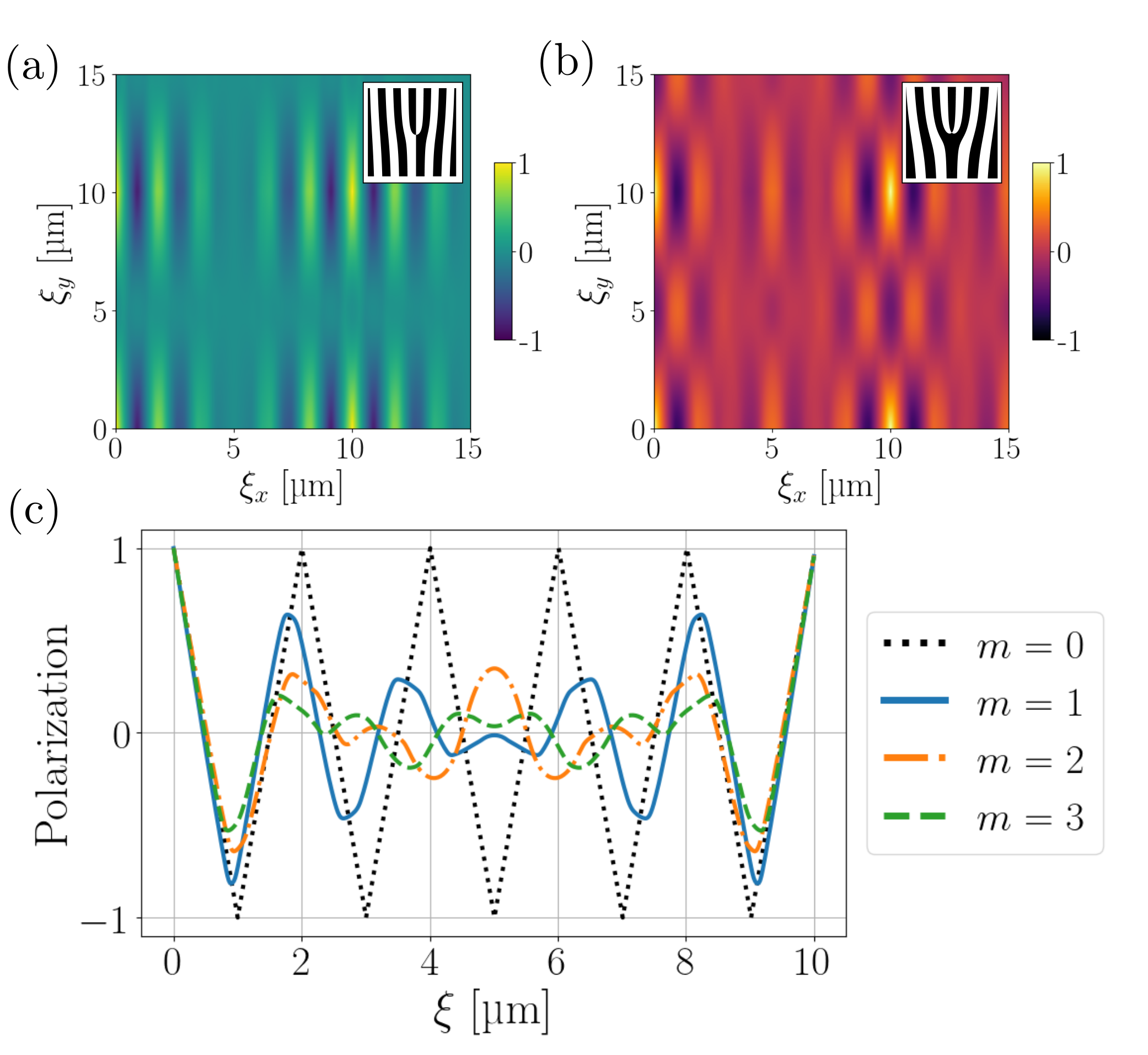}
    \caption{ \label{fig:Charges comparison}
    Simulated SESANS polarization for a plaquette with charge (a) $m=1$ and (b) $m=2$, with the insets being the original FDG profile.
    For visual clarity, we used a constant $\lambda=\SI{0.4}{\nano \meter}$ incident neutron spectrum and a $10\times10$ \SI{}{\micro \meter^2} plaquette size with $d = \SI{38}{\micro \meter}$ (such that $\Phi = -\pi \chi$) and $p = \SI{2}{\micro \meter}$. Notice that the polarization is periodic in both $\xi_x$ and $\xi_y$ with plaquette size.
    These plots demonstrate that SESANS can serve as a direct probe of the sample's topological structure as the polarization signal for each $m$ is markedly distinct. 
    (c) Slices through the origin along $\xi_y=0$ of the simulated SESANS polarizations for plaquettes of various charges. These particular $\xi = \xi_x$ slices of the 2D polarizations correspond to the case where the grooves in the $\hat y$ direction are perpendicular to the encoding direction $\hat x$.
    An arbitrary slice through $(\xi_x,\xi_y)$ can be chosen by changing the angle between the sample and encoding direction.}
\end{figure}

The phase accumulated by the neutron due to the nuclear interaction as it passes through our quasi-2D silicon FDG is $\Phi(x,y)=-\rho_{\ce{Si}} \lambda d \chi(x,y)$, where $\rho_{\ce{Si}} \approx \SI{2.07e-4}{\nano \meter^{-2}}$ is the scattering length density of silicon and $\lambda$ the neutron wavelength \cite{Supp}.
Assuming plane wave illumination $e^{ikz}$ with $k = 2 \pi/\lambda$, the neutron wavefunction immediately after the FDG is given by $e^{ikz}e^{i\Phi(x,y)}$, with 
\begin{equation} \label{eq:plane wave diffraction}
\begin{aligned}
    e^{i\Phi(x,y)} &= D_0 + \frac{1}{2}\left( e^{-i \rho_c \lambda d} - 1 \right) \sum_{n=1}^{\infty} D_{nm}(\alpha), \\
    D_{nm}(\alpha) &= \sinc\left(n \pi /2\right) \left[ e^{in \alpha} + e^{-i n \alpha}\right],
\end{aligned}
\end{equation}
where $D_0 = \left(e^{-i \rho_c \lambda d} + 1\right) / 2$ is the transmission amplitude of the unscattered beam and $\sinc(x) = \sin(x)/x$.
From eq. \eqref{eq:plane wave diffraction}, we see that the angular deviations of the diffraction orders relative to the transmitted beam are $\theta \approx \pm n\lambda/p$, and each order carries an $\ell =\mp n m$ OAM state.
Due to the $\sinc$ weighting of the diffraction orders, only the odd orders of $n$ contribute, and the $n = 1$ conjugate orders dominate.

We present the full plane wave limit calculation for the diffracted neutron state in the supplemental material \cite{Supp}. We find the far-field analytic form of the Bragg donut intensities shown in Fig. \ref{fig:FDGs and SANS}(b) to be
\begin{equation}
\begin{aligned}
    \left| D_{nm}^{(\infty)} \right|^2 =& \  C_{nm} \left| (R q')^{|nm|} \ {}_1F_2 \left( \begin{matrix} \bm a \\ \bm b \end{matrix} \ ; \frac{- R^2 q'^2}{4}\right) \right|^2, \\
    C_{nm} =& \left|\frac{\pi R^2}{\lambda} \frac{ (e^{-i \rho_c \lambda d} - 1)\ \sinc(n \pi /2)}{2^{|nm|+1}(2 + |nm|)(|nm|)!} \right|^2,
\end{aligned}
\end{equation}
where ${}_1F_2$ is a generalized hypergeometric function with $\bm a = \left(1 + |nm|/2 \right)$ and $\bm b = \left(2 + |nm|/2, 1 + |nm| \right)$ \cite{Olver_2010}, $R>0$ a regulator used to ensure convergence of the scattering amplitude integral, and $q'$ the radial coordinate in reciprocal space relative to the center of the Bragg donut.
The upshot of this result is that each Bragg donut for definite $|nm|$ is a linear combination of Bessel and the related Struve functions and \textit{not} pure Laguerre-Gauss modes \cite{Allen_1992} as has previously been asserted \cite{Sarenac_2022}.
Importantly, our calculation shows that the radius of the Bragg donut (defined as the maximum intensity) scales \textit{linearly} with the topological charge $m$ for $m \geq 1$ \footnote{Numerical simulations sometimes seem to imply that the \textit{mean} of the Bragg donut scales like $\sqrt{m}$ for $m \geq 1$ depending on the chosen plaquette parameters, but this is actually due to the fact that the Bragg donuts begin to overlap for larger $m$, in which case the mean intensity is artificially pushed towards the Bragg donut center by the adjacent diffraction orders.}, and is \textit{independent} of $\Delta_t$ since our plane wave calculation assumes the $\Delta_t \to \infty$ limit.
Similar results were found for the analogous optical near- and far-field diffraction from FDGs \cite{Janicijevic_2008,Topuzoski_2011}.

In principle, these Bragg donuts can be observed using the technique of small-angle neutron scattering (SANS) as was previously attempted \cite{Sarenac_2022,Sarenac_2024}, although those data were difficult to interpret due to both a large wavelength bandwidth and shallow grating grooves.
Instead, we performed a SESANS measurement of our FDGs; contrary to SANS, SESANS is a real-space technique that measures the relative phase accumulated by the neutron spin states as they pass through the sample as an experimentally tunable function of distance \cite{Rekveldt_1996} while SANS is an amplitude-only measurement of the scattered neutron intensity.
In SESANS, the neutron scattering angle in one direction is encoded (i.e., Larmor labeled \cite{Rekveldt_2011}) into the neutron's polarization, and so the scattering cross section can be determined by only measuring the polarization.
SESANS also has the advantage that the observed polarization is independent of the intrinsic neutron coherence width unlike the cross section in SANS \cite{McKay_Irfan2024}.

As shown in Fig. \ref{fig:SESANS setup}, in our SESANS setup, each neutron is initially polarized in the $\hat y$ direction and is then coherently split into two paths by the first radio-frequency (rf) neutron spin flipper; these independent trajectories are made parallel by a second rf flipper.
The neutron state is therefore spin-path entangled in the spin and position degrees of freedom \cite{Shen_2020,Shufan_2020,Kuhn_McKay_2023}.
The \textit{entanglement length} $\xi$ is the real-space distance that the entangled neutron spin states are separated and thus the length scale probed in the sample.
For our setup,
\begin{equation} \label{eq:spin echo length}
    \xi = \frac{2 m_n f L_{\mathrm{RF}} \cot \theta_0}{h} \lambda^2 = \xi_0 \lambda^2,
\end{equation}
where $m_n$ is the neutron mass, $f = \SI{2}{\mega \hertz}$ the rf flipper frequency, $L_{\mathrm{RF}} = \SI{1.2}{\meter}$ the distance between each pair of rf flippers, $\theta_0 = \SI{40}{\degree}$ the rf flipper inclination angle, and $h$ Plank's constant.
With these parameters, the \textit{entanglement constant} was fixed to $\xi_0 \approx \SI{1.37e4}{\nano \meter^{-1}}$ for all measurements reported here.
The neutron wavelength band used was 0.3-\SI{1.05}{\nano \meter}, which translates to an entanglement length range of 1.2-\SI{15.}{\micro \meter}.
After scattering from the sample, the final two rf flippers spatially recombine the neutron spin states for subsequent polarization analysis and detection.

Using the \textit{phase-object approximation} (POA) for the scattering amplitude, we can efficiently simulate the expected SESANS polarization signal from our FDG \cite{Supp}. 
The simulations shown in Fig. \ref{fig:Charges comparison} demonstrate that the polarization strongly depends on the grating's topological charge. This is to be expected as the diffraction pattern depends on the topological charge [c.f. Fig. \ref{fig:FDGs and SANS}(b)], and the SESANS polarization in the plane wave limit is precisely the cosine Fourier transform of the diffraction pattern, projected onto a single axis \cite{McKay_Irfan2024}.

\emph{Results and discussion.}---
We performed the experiment at the Larmor instrument at the ISIS neutron and muon source located in the UK; see Fig. \ref{fig:SESANS setup} for a sketch of the experimental setup.
We measured the FDGs in two orientations, the first with the encoding direction perpendicular to the grooves (the perpendicular orientation) and the second with the encoding direction parallel to the grooves (the parallel orientation).
In reference to Fig. \ref{fig:Charges comparison}, in general the entanglement length is given by $\xi = \sqrt{\xi_x^2 + \xi_y^2}$. With our convention, the perpendicular orientation corresponds to $\xi = \xi_x$ and the parallel orientation to $\xi = \xi_y$.

\begin{figure}[tb!]
    \centering
    \includegraphics[width=0.95\linewidth]{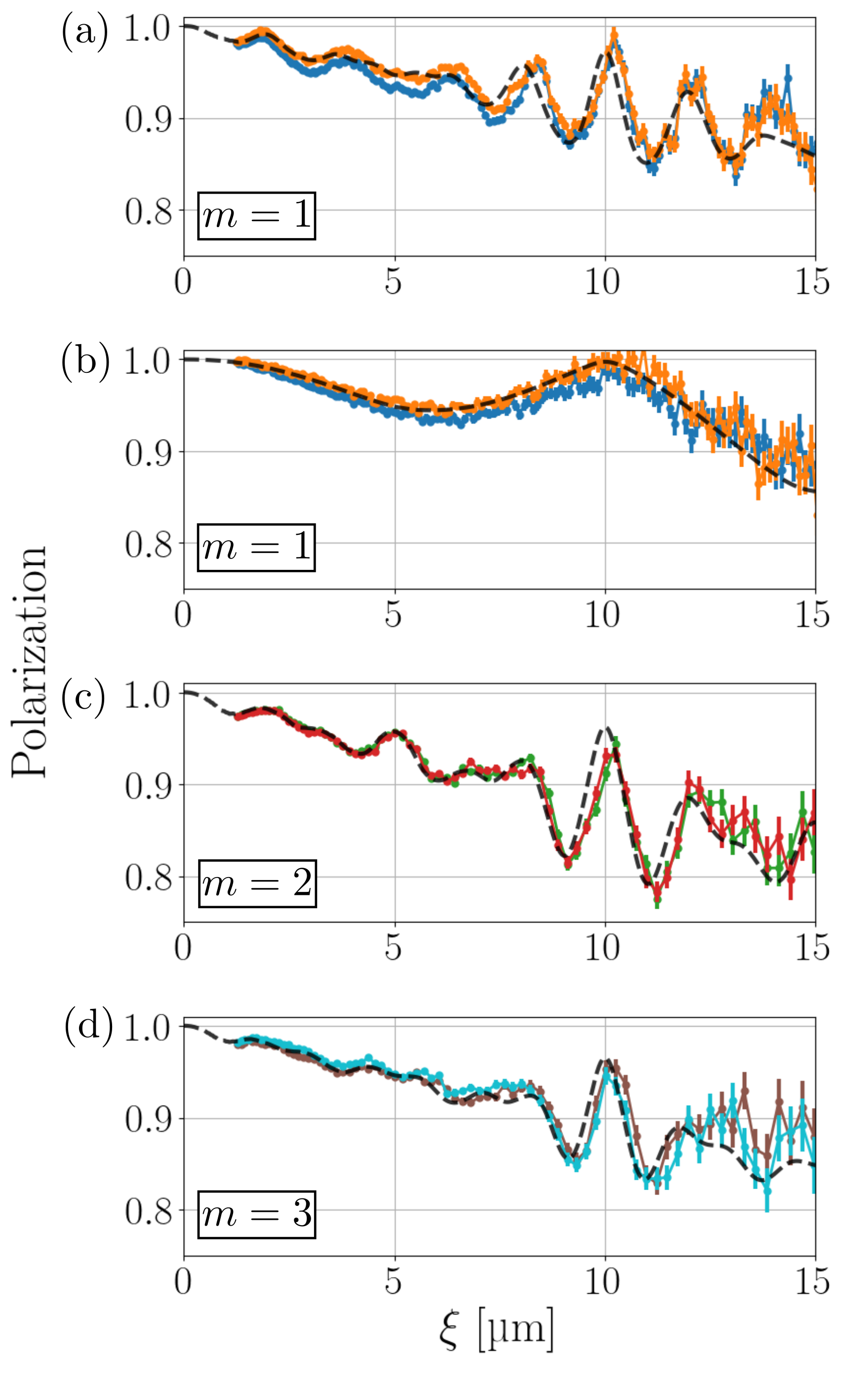}
    \caption{ \label{fig:Exp data}
    Measured SESANS polarization data of two charge $m=1$ gratings in the (a) perpendicular and (b) parallel orientation.
    Measured polarizations of two charge (c) $m=2$ and (d) $m=3$ gratings in the perpendicular orientation.
    In all parts, the dashed black trace corresponds to a SESANS POA simulation convolved with the instrument resolution function discussed in the supplemental material \cite{Supp}. The two solid traces and data points in each part correspond to the measured polarizations from a particular FDG. }
\end{figure}

\begin{figure}[tb!]
    \centering
    \includegraphics[width=0.95\linewidth]{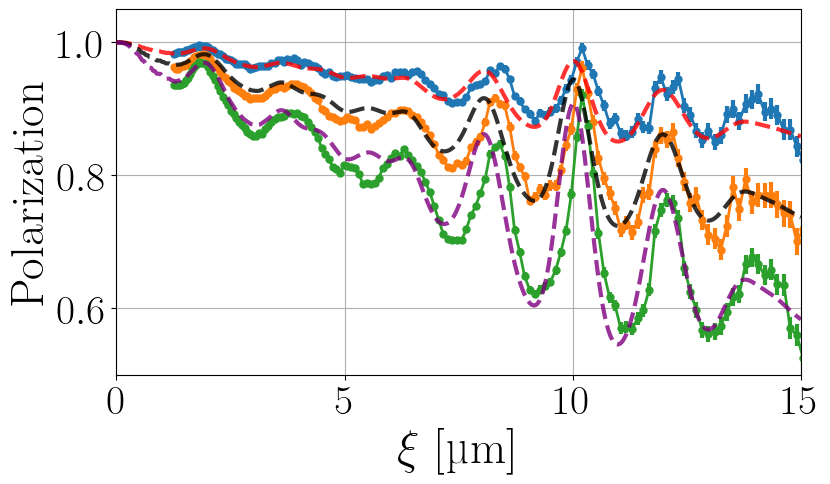}
    \caption{ \label{fig:Exp data stacked}
    Comparison of the measured SESANS polarization data of a single grating (blue), and two (orange) and three (green) stacked charge $m=1$ gratings.
    The red dashed trace is the simulation result for a single charge $m=1$ grating of depth $d=\SI{5.5}{\micro \meter}$ reproduced from Fig. \ref{fig:Exp data}(a), and the black and purple dashed traces the result of a simulation with a single FDG of depth $d\sqrt{2}$ and $d \sqrt{3.5}$, respectively. For the three-grating simulation, we attribute the deviation from the expected apparent depth of $d\sqrt{3}$ to the fact that the third grating was produced by a slightly different process that most likely resulted in slightly deeper grooves \cite{Supp}. }
\end{figure}

As shown in Fig. \ref{fig:Exp data}, there is good agreement between the data (solid traces) and the POA simulations (dashed traces).
The sloped background seen in Figs. \ref{fig:Exp data} and \ref{fig:Exp data stacked} is an artifact of using a pulsed time-of-flight neutron source [see eqs. \eqref{eq:plane wave diffraction} and \eqref{eq:spin echo length}].
We measured a total of six gratings individually in the perpendicular orientation, two each of charge 1, 2, and 3.
The intricate pattern seen in each of the SESANS polarizations serves as the unique, identifying fingerprint of the particular OAM state.
We also present the measured polarization from two charge $m=1$ gratings in the parallel orientation in Fig. \ref{fig:Exp data}(b); the observed polarization in this orientation is relatively featureless and similar for all charges, only displaying a single broad peak at \SI{10}{\micro \meter} corresponding to the inter-plaquette spacing.
For each of the POA simulations shown in Figs. \ref{fig:Exp data} and \ref{fig:Exp data stacked}, the only free parameter was the depth of the grooves which was varied between 3-6 microns in order to match the experimentally observed polarization contrast.
Specifically, the charge 1, 2, and 3 data were best fit using a depth of $\SI{5.5}{\micro \meter}$, $\SI{3.9}{\micro \meter}$, and $\SI{4.2}{\micro \meter}$, respectively.
These SESANS-determined depths are consistent with the depths found from SEM images shown in the supplemental material~\cite{Supp}.

To enhance the polarization contrast, we also performed measurements with FDGs stacked back-to-back, with gratings separated from one another by $\sim \! \SI{1}{\milli \meter}$. As shown in Fig. \ref{fig:Exp data stacked}, the observed polarization from the stacked gratings is the product of the single-grating data.
Therefore, stacking gratings of the same charge and groove depth $d$ results in a similar polarization to that which would be found from a single grating with an increased groove depth of $d \sqrt{n_g}$, where $n_g$ is the number of gratings.
Importantly, the individual plaquettes were not required to be placed in registry to observe this enhancement.
Because etching FDGs with a depth-period aspect ratio greater than $\sim \! 5:1$ is challenging, stacking multiple gratings can provide a simpler alternative means to enhance the contrast.

\emph{Conclusion.}---
We have demonstrated the production of neutron OAM beams with forked phase gratings using the SESANS technique.
The experiments presented here also show that SESANS is extremely sensitive to real-space structures, being able to discriminate between samples with similar but ultimately
distinct features, and so can be used for example as a direct probe of topological defects in materials.
We attribute this sensitivity to the fundamental fact that SESANS is a phase-sensitive, interferometric technique.

We suspect that these OAM neutron beams will be effective probes of both the structure \cite{Mallick_2024} and dynamics \cite{Fujita_2017} of topological systems such as skyrmions \cite{Henderson_2024} and other magnetically textured materials \cite{Göbel_2021} that have spintronic applications \cite{Kang_2016,Zhang_2020,Petrović_2025}.
Additionally, spin-orbit neutrons beams that are entangled in their spin and OAM degrees of freedom are another promising type of OAM beam for the measurement of topological materials \cite{LeThien_2023}.
On the other hand, grating interferometry with FDGs has been proposed in order to both enhance the imaging resolution as well as serve as a direct probe of chiral structures \cite{Zhou_2021,Dettlaff_2023}.

Finally, we acknowledge that the exact nature of the OAM content of our neutron beam remains an open question, namely whether this OAM is truly a single-neutron or only a beam property. Elucidation of this point requires both higher-precision measurements of the intrinsic coherence width of the neutron as well as a more sophisticated theoretical analysis that ab initio incorporates the neutron's intrinsic coherence width.

\FloatBarrier
\section*{Acknowledgements}
\begin{acknowledgments}

The grating fabrication was conducted as part of a user project at the Center for Nanophase Materials Sciences (CNMS), which is a US Department of Energy, Office of Science User Facility at Oak Ridge National Laboratory.
The experiment was performed on the Larmor instrument at the ISIS Neutron and Muon source (UK) supported by beamtime allocations RB2410229 and RB2510242 from the Science and Technology Facilities Council \cite{UK_beamtime1,UK_beamtime2}.
The IU Quantum Science and Engineering Center is supported by the Office of the IU Bloomington Vice Provost for Research through its Emerging Areas of Research program.
This work was funded in part by the Department of Energy STTR program (grant no. DE-SC0023624).

\end{acknowledgments}

\FloatBarrier
\bibliographystyle{apsrev4-2.bst}
\bibliography{sources.bib}

\clearpage
\newpage
\section*{Supplemental material}

\subsection{FDG etch procedure and SEM images}

\begin{figure}[h!]
    \centering
    \includegraphics[width=0.8\linewidth]{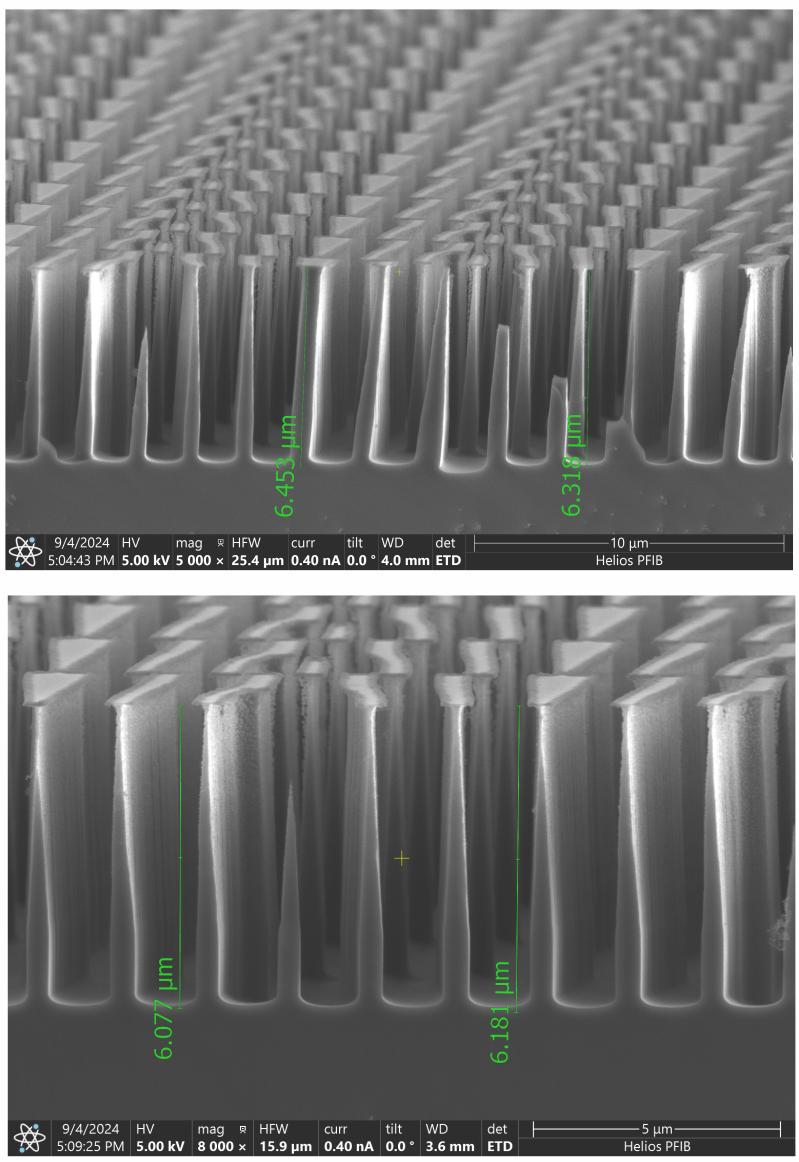}
    \caption{ \label{fig:FDG SEM}
    Side-view SEM images of first set of charge $m=1$ FDGs. Notice that although the depth is roughly \SI{6}{\micro \meter}, there is also significant undercutting of the walls near the top of grating, which results in a trapezoidal or even triangular profile rather than the desired rectangular profile. The undercutting could also reduce the uniformity of the plaquettes if it is not constant over the entire etched portion of the wafer.
    The combination of these two effects reduce the measured SESANS polarization contrast from the simulated contrast [see Fig. \ref{fig:Exp data}(a) in the main text]. }
\end{figure}

The charge $m=1$ binary silicon forked phase gratings were produced at the Center for Nanophase Materials Sciences (CNMS) at Oak Ridge National Laboratory (ORNL). Two different etching processes were developed, the first of which is as follows: an atomic layer deposition of \SI{70}{\nano \meter} of \ce{Al_2O_3} was performed on a \ce{Si} wafer substrate. Then, the wafer was spin-coated with ZEP520 electron-beam photoresist at 1 kRPM followed by a baking at 180 \degree C for 2 minutes. Next, the wafer was exposed to a lithographic electron beam with current \SI{10}{\nano \ampere}, with a total exposure time of 12-13 hours per $10 \times \SI{10}{\milli \meter^2}$.
The wafer was then developed with a 1 minute xylene wash, and the remaining alumina was sputter etched away in \ce{Ar} plasma for 3 minutes.
Finally, a 10 minute reactive ion etch of the \ce{Si} in \ce{SF_6}/\ce{C_4F_8} was performed with a 20 minute \ce{O_2} chamber clean. This ion etch was repeated 3 times.
The second process was mostly equivalent, except that \SI{15}{\nano \meter} of \ce{Cr} was used instead of the \ce{Al_2O_3}, and the final ion etch was repeated 4 times.

\begin{figure}[t!]
    \centering
    \includegraphics[width=0.8\linewidth]{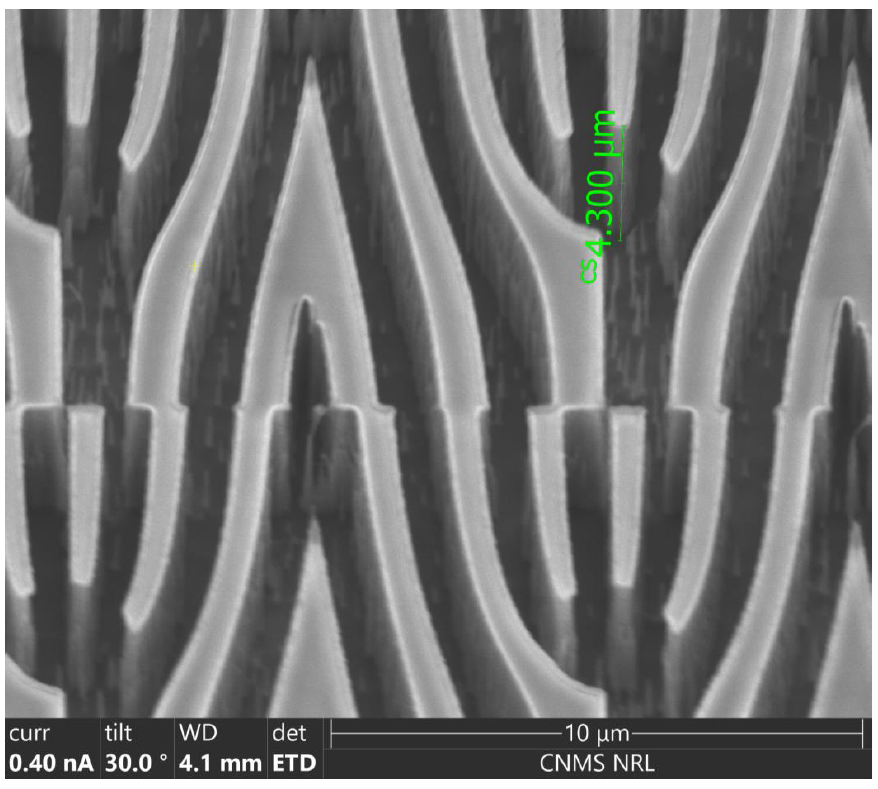}
    \caption{ \label{fig:FDG SEM new}
    Example SEM image of a charge $m=3$ FDG using the improved process. Notice that the groove profile is much more rectangular than the $m=1$ grating profile shown in Fig. \ref{fig:FDG SEM}. }
\end{figure}

As shown in Fig. \ref{fig:FDG SEM}, there appears to be significant undercutting of walls of the grating, most likely due to thermal gradients present in the wafer during the etching process. 
To account for this in our simulations, note that a 50\% duty cycle triangular-profile grating can be described by the indicator function
\begin{equation}
    \chi_{\mathrm{tri}}(x,y) = \max\left[0, 1 - \frac{2}{\pi} \arccos \left( \cos \alpha \right) \right],
\end{equation}
where again $\alpha = (2\pi/p) x - m \phi$ as in the main text. Similarly, a 50\% duty cycle trapezoidal indicator function can be expressed as
\begin{equation}
    \chi_{\mathrm{trap}}(x,y) = \min\left[1, c \, \chi_{\mathrm{tri}}(x,y) \right],
\end{equation}
where the parameter $c \geq 1$ determines the width of the plateau ($c=1$ gives a triangular profile with zero plateau width).
For simplicity, we used the triangular indicator function for our simulations of the $m=1$ grating shown in Figs. \ref{fig:Exp data} and \ref{fig:Exp data stacked} in the main text. 

To limit the undercutting observed in the $m = 1$ FDGs, we modified the etching process for the $m=2$ and $m=3$ gratings. 
As the processes were similar to the $m=1$ processes, we only highlight the differences:
only \SI{50}{\nano \meter} of \ce{Al_2O_3} were deposited on the \ce{Si} wafer substrate, the spin-coating was performed at 1.5 kRPM, the xylene wash was limited 40 seconds, the \ce{Ar} sputter etch was limited to 2 minutes, and the reactive ion etch was only performed twice, the first for 11 minutes and the second for 6 minutes.
The second modified process was similar with \SI{15}{\nano \meter} of \ce{Cr} was used instead of the \ce{Al_2O_3}, and the final ion etch was performed twice for 11 minutes each.

\subsection{The POA and SESANS}

The phase-object approximation (POA) is a common method used in both the analysis of SANS and SESANS experiments that can account for multiple scattering as long as each individual scattering is weak \cite{Weiss_1951,Glauber_1959,deHaan_2007}.
The POA assumes that the neutron wavefunction can be well-described by the WKB approximation during the scattering process; the resulting neutron state is the coherent combination of the scattering state from the well-known eikonal approximation that is treated in most scattering textbooks \cite{Gottfried_2003,Sakurai_2021} along with a contribution of the transmitted (i.e., unscattered) beam.
The POA elastic cross section is given by $d \sigma / d \Omega = |f_{\mathrm{po}}(\bm q)|^2$, with elastic scattering amplitude 
\begin{subequations} \label{eq: po total}
\begin{align} 
    f_{\mathrm{po}}(\bm q) &= \frac{-i}{\lambda} \int_{A} d\bm r_{\perp} \, e^{-i \bm q \cdot \bm r_{\perp}} e^{i \Phi(\bm r_{\perp})}, \label{eq:po_general} \\
    \Phi(\bm r_{\perp}) &= \frac{-m_n \lambda}{2 \pi \hbar^2} \int_{s} d r_{\|} \, V(\bm r). \label{eq:phase_function}
\end{align}
\end{subequations}
In these equations, $\bm r = \bm r_{\perp} \oplus r_{\|}$, the integral over $\bm r_{\perp}$ is taken over the entire illuminated area $A$ of the sample, $m_n$ is the mass of the neutron, $s$ the classical path of neutron through the sample, and $\bm q = \bm k_i - \bm k_f$ is the momentum transfer.
The fact that the above scattering amplitude and cross section also contain the contribution from the transmitted beam can be seen by considering that the limit $V \to 0$ in eq. \eqref{eq: po total} is non-zero and proportional to the beam size.
Finally, notice that the POA reduces to the first Born approximation plus a contribution from the transmitted beam when $e^{i\Phi(\bm r_{\perp})}$ is expanded to first order.

For quasi-2D gratings fixed in the $x$-$y$ plane with plane wave illumination $e^{i k z}$, $\Phi(\bm r_{\perp})$ simplifies to
\begin{subequations}
\begin{align}
    \Phi(x,y) &= -\lambda \int_{\mathbb{R}} dz' \, V(x,y,z') = -\rho_c \lambda d \,\chi(x,y), \\
    \chi(x,y) &= \begin{cases} 1 \text{ if } (x,y) \in A_g,\\ 0 \text{ otherwise}, \end{cases}
\end{align}
\end{subequations}
where $\rho_c$ is the coherent scattering length density (SLD), $\lambda$ the neutron wavelength, $d$ the depth of the grooves, and $A_g$ the set of points that mathematically define the region containing the grooves. 
This form assumes that the nuclear scattering can be modeled using the Fermi pseudopotential \cite{Squires_2012}, in which case $V(\bm r) \approx (2 \pi \hbar^2/m_n) \rho_c(\bm r)$ and also neglects the imaginary part of the SLD as the thermal and cold neutron absorption of most elements is weak.
Notice that the scattering amplitude neatly separates into two components, namely a \textit{contrast} term given by $-\rho_c \lambda d$ and an \textit{indicator} term $\chi$ that determines the shape of the scattering signal.
As discussed in the main text, for our gratings,
\begin{equation}
    \Phi(x,y) = -\rho_{\ce{Si}} \lambda d \chi(x,y),
\end{equation}
where $\rho_{\ce{Si}} \approx \SI{2.07e-4}{\nano \meter^{-2}}$ is the SLD of \ce{Si} with indicator function given in eq. \eqref{eq:Rect FDG} in the main text and repeated below for convenience:
\begin{equation}
    \chi(x,y) = \frac{1}{2} \left[ 1 + \sign \left( \cos \alpha \right)\right],
\end{equation}
with $\alpha = (2\pi/p) x - m \phi$, where $p$ is the period, $m \in \mathbb{Z}$ the topological charge, and $\phi = \arctan(y/x)$.

Next, we discuss the derivation of eq. \eqref{eq:plane wave diffraction} in the main text. Expanding $\chi$ as a Fourier series gives
\begin{equation} \label{eq:chi fourier}
    \chi(x,y) = 1/2 + \sum_{n=1}^{\infty} \sinc(n \pi /2) \cos(n \alpha).
\end{equation}
Equation \eqref{eq:plane wave diffraction} in the main text is then found directly from the relation
\begin{equation} \label{eq:thingy}
    e^{-i\rho_c \lambda d \chi(x,y)} = e^{-i\rho\lambda d} \, \chi(x,y) + 1-\chi(x,y),
\end{equation}
which follows from $\chi \in \{0,1\}$.
Combining eqs. \eqref{eq:chi fourier} and \eqref{eq:thingy}, we see that the amplitude of the transmitted beam is 
\begin{equation}
    D_0 = \frac{1}{2}\left( e^{-i \rho_c \lambda d} + 1 \right).
\end{equation}
Notice that for a $\pi$-shift phase grating, $D_0 = 0$ and so there is no transmitted beam.

Finally, we provide the connection between the POA cross section and the measured SESANS polarization. Assuming that the incident neutron state is a plane wave, the scattering angle is small, the detector captures all of the scattering, and $\hat x$ is the encoding direction, the empty-beam normalized polarization $P_0$ is given by
\begin{align} \label{eq:SESANS POA}
    \frac{P(\xi)}{P_0} &= \frac{1}{\sigma} \int_{\mathbb{R}^2} d q_x d q_y \, \frac{d \sigma}{d \Omega}(q_x, q_y) \cos(q_x \xi) \nonumber \\
    &= \frac{1}{\sigma} \Re \left[ \int_{\mathbb{R}^2} d x d y \,  e^{i \Phi(x + \xi,y)} e^{-i \Phi(x,y)} \right],
\end{align}
where $\sigma$ is the total scattering cross section and $\Re[\cdot]$ denotes the real part. From the final line eq. \eqref{eq:SESANS POA}, we see that the SESANS polarization directly measures the real-space correlation for all points separated by the entanglement length $\xi$.
Notice that from the projection-slice theorem, eq. \eqref{eq:SESANS POA} can be recast as a Fourier transform in $q_x$ followed by a slice through $q_y = 0$.
This procedure is generally the most computationally efficient method of calculating the expected SESANS polarization rather than by directly computing the autocorrelation function.

\subsection{Analytical solution of FDG plane wave diffraction}

In this section, we derive the analytical solution of the diffracted OAM state from plane wave illumination in the POA in close analogy to Brand \cite{Brand_1999}.
From eq. \eqref{eq:Rect FDG} in the main text and eq. \eqref{eq: po total}, we see that 
\begin{equation} \label{eq: po f amp}
\begin{aligned}
    f_{\mathrm{po}} =& \frac{-i}{\lambda} \frac{e^{-i \rho_c \lambda d} - 1}{2} \sum_{n=1}^{\infty} \sinc(n \pi /2) \times \\
    &\int_A r dr d\phi \, e^{-i q_x r \cos \phi} e^{-i q_y r \sin \phi} \cos(n \alpha),
\end{aligned}
\end{equation}
where we transformed into polar coordinates $(r\cos \phi,r\sin \phi)$ and neglected the contribution to $f_{\mathrm{po}}$ from the transmitted beam.
Next, to perform the integral over $\phi$, we use the Jacobi-Anger expansion 
\begin{equation}
    e^{i \alpha r \cos \phi} = \sum_{l=-\infty}^{\infty} i^l J_{l}(\alpha r) e^{i l \phi},
\end{equation}
where $J_l$ are Bessel functions of the first kind and the relation $a \cos \phi + b \sin \phi = \sqrt{a^2 + b^2} \cos(\phi - \beta)$ with $\beta = \arctan (b/a)$.
Doing so, we find that the amplitude for the $n\textsuperscript{th}$ diffraction order is 
\begin{equation}
\begin{aligned}
    f^{(n)}_{\mathrm{po}} \propto \sum_{l=-\infty}^{\infty} i^l \int_A r dr d\phi \, \Big[ & J_l(\alpha_- \, r) e^{i(\phi - \beta_-)l}e^{-inm\phi} + \\
    & J_{l}(\alpha_+ \, r) e^{i(\phi - \beta_+)l}e^{inm\phi} \Big]
\end{aligned}
\end{equation}
where we defined the parameters
\begin{subequations}
\begin{align}
    \alpha_{\pm} =& \sqrt{(q_x \pm 2\pi n/p)^2 + q_y^2}, \\
    \tan \beta_{\pm} =& q_y / (q_x \pm 2\pi n/p).
\end{align}
\end{subequations}
Notice that $\alpha_{\pm}$ and $\beta_{\pm}$ depend on the order $n$.

Because $\int_0^{2\pi} d\phi \, e^{i\phi(l \mp nm)} = 2\pi \delta_{l,\pm nm}$, the total scattering amplitude becomes
\begin{equation} \label{eq:POA FDG}
\begin{aligned}
    f_{\mathrm{po}} = \sum_{n=1}^{\infty} A_n \int r dr \, \Big[&  i^{nm} J_{nm}(\alpha_- \, r) e^{-i\beta_-nm} + \\
    & i^{-nm} J_{-nm}(\alpha_+ \, r) e^{i\beta_+nm} \Big] ,
\end{aligned}
\end{equation}
where the proportionality factor is given by
\begin{equation}
    A_n = \frac{-i \pi}{\lambda} \frac{e^{-i \rho_c \lambda d} - 1}{2} \sinc(n \pi /2).
\end{equation}
At this point, we can see that the intensities at the center of each Bragg donut for a particular order vanish by plugging in $q_x = \pm 2\pi n/p$ and $q_y = 0$ into eq. \eqref{eq:POA FDG} and noting that $J_{l\neq0}(0) = 0$; the other orders in the sum will also be highly oscillating at that point, and will contribute only a small background due to the remaining integral over $r$ as we have numerically confirmed.
Next, if we consider an expansion about the center of the Bragg donuts of the form
\begin{equation} \label{eq: radial expansion}
\begin{aligned}
    q_x =& q'\cos\phi' \pm 2\pi n/p, \\
    q_y =& q'\sin\phi',
\end{aligned}
\end{equation}
for $q' \geq 0$ and $\phi' \in [0, 2 \pi)$, we see that the neutron phase is given by $e^{\mp inm \phi'}$, which is indeed the promised result.

For all $q_x$ and $q_y$ sufficiently far away from the diffraction orders, we can perform the integral over $r$ in eq. \eqref{eq:POA FDG} by introducing the regulator $e^{- \epsilon r}$ with $\epsilon> 0$, in which case we obtain
\begin{equation}
\begin{aligned}
    f_{\mathrm{po}} =& \sum_{n=1}^{\infty} nm A_n \left( i^{nm}\frac{e^{-i\beta_-nm}}{\alpha_-^2} + i^{-nm} \frac{e^{i\beta_+nm}}{\alpha_+^2} \right),
\end{aligned}
\end{equation}
where we require $\alpha_{\pm} > 0$ such that the limit $\epsilon \to 0^+$ is valid.
However, as discussed above, the integral must be finite when $\alpha_{\pm} \to 0$ which corresponds to the center of each Bragg donut.
To accomplish this analytically, notice that only one term in the sum will approach zero at each singularity; focusing on that one term in the sum, we can break that integral over $r$ into two parts at the point $r=R>0$.
Note that if $R \gtrsim \pi/p$ as is the case for large $|nm|$, then our estimate fails as the Bragg donuts begin to overlap.
Expanding using eq. \eqref{eq: radial expansion} and neglecting the small contributions from $r>R$ as well as from the conjugate singularity of the same order, we now have
\begin{equation}
\begin{aligned} \label{eq: fpon integral}
    f^{(n)}_{\mathrm{po}} \approx & i^{\pm nm} A_n  e^{\mp i nm \phi'}\int_0^R r dr \, J_{\pm nm}(q' \, r),
\end{aligned}
\end{equation}
which does have a closed form in terms of the generalized hypergeometric function ${}_1F_2$. Using the notation given in \cite{Olver_2010} and fixing $m>0$, we find that
\begin{subequations} \label{eq: final Bragg donut total}
\begin{align} \label{eq: final Bragg donut}
    f^{(n)}_{\mathrm{po}} \approx& B_{nm} (R q')^{|nm|} {}_1F_2 \left( \begin{matrix} \bm a \\ \bm b \end{matrix} \ ; \frac{- R^2 q'^2}{4}\right), \\
    B_{nm} =& \frac{(\pm 1)^{nm} i^{\pm nm} A_n R^2 }{2^{|nm|}(2 + |nm|)(|nm|)!} e^{\mp i nm \phi'},
\end{align}
\end{subequations}
where $\bm a = \left(1 + |nm|/2 \right)$ and $\bm b = \left(2 + |nm|/2, 1 + |nm| \right)$.
The result when $m<0$ can similarly be found by using the relation $J_{-|nm|} = (-1)^{|nm|} J_{|nm|}$.

\begin{figure}[tb]
    \centering
    \includegraphics[width=0.99\linewidth]{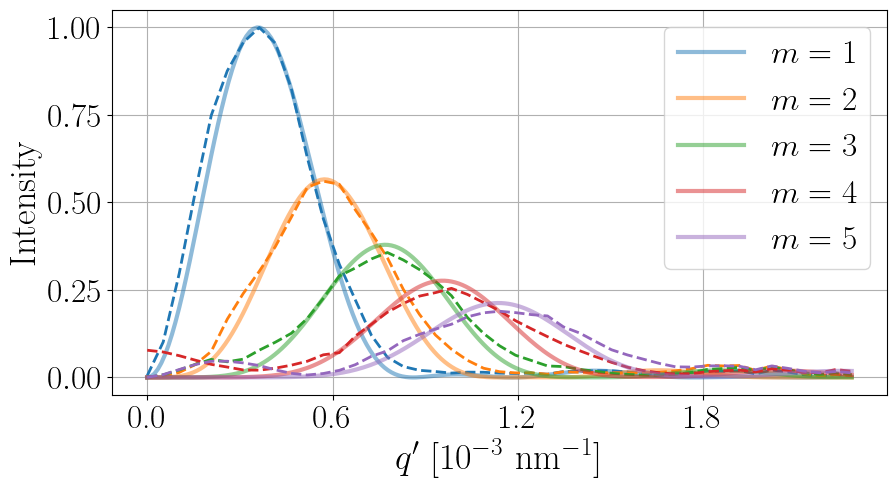}
    \caption{ \label{fig:donut comparison}
    Comparison of the intensities as a function of the radius $q'$ [see eq. \eqref{eq: radial expansion}] about the positive $n=1$ diffraction order calculated using eq. \eqref{eq: final Bragg donut total} (solid traces) and from numerical SESANS POA simulations (dashed traces). Both the $m>1$ numerical and analytical results were independently scaled to the maximum of the $m=1$ peak of the numerical and analytical result, respectively. We also chose $R = 0.4$ to overlap the $m=1$ simulation and analytical solutions. Note that even though the numerical simulations contain additional contributions from all of the other diffraction orders, they still show good agreement with the analytical result that only considers a single order.
    These simulations used \SI{2}{\micro \meter}-period, \SI{5}{\micro \meter}-depth, $10 \times \SI{10}{\micro \meter}^2$ plaquettes.}
\end{figure}

We plot some examples of the intensity profiles generated by these scattering amplitudes in Fig. \ref{fig:donut comparison}. 
We note that these solutions for the Bragg donut profiles are qualitatively similar to but definitely distinct to the solutions consisting of pure Laguerre-Gauss (LG) modes as is commonly asserted in the literature \cite{Heckenberg_1992a,Lee_2019,Sarenac_2022}.
This discrepancy between our and the commonly asserted solution is due to the fact that the conventional solution only considers the state immediately after the sample and erroneously ignores the fact that diffraction changes the spatial profile of the state.
Interestingly, although diffraction does not preserve the spatial structure of the wavefunction, the OAM quantum number $m$ imparted by the topological grating to the neutron is preserved through the diffraction process into the far-field.
Coming back to our solution, we can determine from it that the radius of each Bragg donut defined as the maximum radial value grows \textit{linearly} with topological charge $m$ for $m \geq 1$.
This scaling law has also been confirmed numerically up to $m \sim \! 20$ for a variety of plaquette shapes and periods.
Note that this linear scaling disagrees with the previously asserted scaling law of $\sqrt m$ in \cite{Sarenac_2022} that follows from the assumption that the Bragg donuts are pure LG modes.

We see that at least for the $m$-value range investigated here, the Bragg donut radii for the FDG states grow with $|nm|$ at a noticeably quicker rate than LG modes. Generally, the upper limit of $|nm|$ is determined as the point when the neighboring far-field Bragg donuts become tangent; this criterion determines the maximum useful FDG topological charge.
The so-called ``perfect vortex beams'' were developed to remove this dependence between the Bragg donut radius and the OAM state \cite{Ostrovsky_2013,Yu_2023}. However, it was later shown that this square-root dependence between the Bragg donut radius and the OAM state appears to be a universal property of Bessel-Gauss beams (i.e., circularly symmetric, paraxial vortex beams), and this supposed independence is actually still a square-root dependence but with a very small proportionality coefficient \cite{Pinnell_2019}.
Interestingly, the smallest vortex beam radius for a particular $m$ is given by the simple relation $r_{\mathrm{min}} = m \lambda / (2 \pi)$ \cite{Roux_2003}. Attempting to localize a vortex beam below this size will generate non-propagating evanescent waves that will deplete the beam.

As an aside, we see from eq. \eqref{eq: final Bragg donut total} that the regulator $R$ appears to act like a simple scaling parameter, ultimately determining the radius of the Bragg donut in reciprocal space.
In this light, it is tempting to view the introduction of $R$ as a naive way to model the effect of a finite intrinsic coherence width $\Delta_t$. However, this idea breaks down when one considers that $R$ is only used to prevent an ultraviolet divergence in the integral defined in eq. \eqref{eq: fpon integral} due to slow $r^{-1/2}$ convergence of the integrand to zero \cite{Olver_2010}.
This situation is analogous to the well-known Coulomb potential divergence in standard quantum scattering theory \cite{Gottfried_2003} in which there is always a divergent forward scattering amplitude (in our case, the divergent amplitude appears at the Bragg donut center rather than the forward direction).
While the particular choice of $R$ is somewhat arbitrary as long as $R$ is greater than the position of the first zero of the Bessel function in the integrand of eq. \eqref{eq: fpon integral}, the choice of $R$ has \textit{no} relation to the intrinsic coherence width of the neutron which does not enter to our plane wave calculation.
To include the effects of a finite intrinsic coherence width, one must instead ab initio modify the scattering amplitude of eq. \eqref{eq: po total} to explicitly include $\Delta_t$ as has been explored for example in \cite{Karlovets_2015}, although this is beyond the scope of the present work.
Finally, we note that the use of the regulator $R$ is unnecessary to model a real experiment with finite-sized FDGs. We chose to introduce $R$ rather than assume a finite FDG in order to remove the artifacts due to the choice of boundary shape. For example, a square boundary introduces the standard $\sinc$-like artifacts [e.g., see Fig. \ref{fig:FDGs and SANS}(b) in the main text] and a circular boundary Bessel-like artifacts.
These artifacts are well-understood from the textbook diffraction theory \cite{Born_Wolf_2005}, and so we do not consider them here.

Coming back to the solution given in eq. \eqref{eq: final Bragg donut total}, we can gain some insight into this complicated result by instead applying the well-known Bessel function recursive identity 
\begin{equation}
    rJ_{|nm|} = 2(|nm| -1) J_{|nm|-1} - r J_{|nm|-2}
\end{equation}
directly to the integrand in eq. \eqref{eq: fpon integral}.
From this form, we see that the integral for a definite $|nm|$ reduces to a linear combination of $J_0$ and $J_1$ with polynomial coefficients in $R q'$ when $nm \in 2 \mathbb{Z}$ and a similar linear combination of $J_0 H_1$ and $J_1 H_0$ where $H_0$ and $H_1$ are the zeroth- and first-order Struve functions when $nm \in 2 \mathbb{Z} + 1$ \cite{Olver_2010}.

Finally, to connect the results presented in this section to the diffraction angle, note that for elastic small-angle scattering, $q \approx k \theta$, and so indeed the angular deviations of the diffraction orders from the transmitted beam are $\theta \approx \pm n\lambda/p$ as discussed in the main text.

\subsection{SESANS data treatment}

The instrument resolution function was taken to be Gaussian for simplicity. The wavelength spread $\delta \lambda$ at the Larmor instrument can be well-approximated as linear in $\lambda$ as discussed in \cite{McKay_Irfan2024}, in which case the entanglement length uncertainty becomes approximately proportional to $\xi$ as $\delta \lambda$ is dominated by the beam divergence standard deviation $\delta \theta_0 \approx \SI{10}{\milli \radian}$.
Therefore, the $\xi$-dependent standard deviations of our Gaussian resolution function are $\delta \xi / \xi \approx \SI{2.0e-2}{}$ as determined by a linear fit.

To reduce the statistical error of point-by-point division, we fitted the empty beam polarization $P_0$ to a generic 5\textsuperscript{th} order Chebyshev polynomial. The fitted coefficients varied slightly between the data sets, and we attribute these small variations to thermal fluctuations and magnetic field contamination.

\end{document}